\documentclass[reprint,preprintnumbers,nofootinbib,superscriptaddress,amsmath,amssymb,aps,prd]{revtex4-2}
\pdfoutput=1
\usepackage{graphicx,hyperref,braket,orcidlink, algorithm2e, amssymb, amsmath, dsfont,subfigure,multirow}
\usepackage{lineno}
\graphicspath{{figs/}}
\hypersetup{
	unicode=true,
	pdftoolbar=true,
	pdfmenubar=true,
	pdffitwindow=false,
	pdfstartview={FitH},
	pdftitle={syk},
	pdfauthor={J.Y.~Araz, R.G.~Jha, F.~Ringer, B.~Sambasivam}, 
	pdfsubject={syk},
	pdfcreator={J.Y.~Araz, R.G.~Jha, F.~Ringer, B.~Sambasivam},
	pdfproducer={}, 
	pdfkeywords={},
	pdfnewwindow=true,
	colorlinks=true,
	linkcolor=blue,
	citecolor=magenta,
	filecolor=magenta,
	urlcolor=blue
}

\definecolor{pc1}{rgb}{0.69, 0.25, 0.21}
\newcommand{\CHANGE}[1]{\textcolor{black}{{#1}}}

\usepackage{braket}
\usepackage{dsfont}
\usepackage{nccmath}
\usepackage{scalerel}
\usepackage{bm}
\usepackage{xcolor}
\definecolor{rindou1}{rgb}{0.4431,0.2862,0.7960}
\definecolor{rindou2}{rgb}{0.0078,0.1215,0.4392}
\definecolor{lapis}{rgb}{0.0.0470,0.2941,0.5568}
\definecolor{mn}{rgb}{0.15, 0.35, 0.95}

\begin{document}
\title{Thermal state preparation of the SYK model using a variational quantum algorithm} 

\author{Jack Y. Araz\orcidlink{0000-0001-8721-8042}}
\email{jackaraz@jlab.org}
\affiliation{Thomas Jefferson National Accelerator Facility, Newport News, VA 23606, USA}
\affiliation{Department of Physics, Old Dominion University, Norfolk, VA 23529, USA}
\author{Raghav G. Jha\orcidlink{0000-0003-2933-0102}}
\email{raghav.govind.jha@gmail.com}
\affiliation{Thomas Jefferson National Accelerator Facility, Newport News, VA 23606, USA}
\author{Felix Ringer\orcidlink{0000-0002-5939-3510}}
\email{fmringer@jlab.org}
\affiliation{Thomas Jefferson National Accelerator Facility, Newport News, VA 23606, USA}
\affiliation{Department of Physics, Old Dominion University, Norfolk, VA 23529, USA}
\affiliation{Center for Nuclear Theory, Department of Physics and Astronomy,
Stony Brook University, NY 11794, USA}
\author{Bharath Sambasivam\orcidlink{0000-0002-5765-9469}}
\email{bsambasi@syr.edu}
\affiliation{Department of Physics, Syracuse University, Syracuse, NY 13244, USA}

\preprint{JLAB-THY-24-4088}

\begin{abstract}
We study the preparation of thermal states of the dense and sparse Sachdev-Ye-Kitaev (SYK) model using a variational quantum algorithm for $6 \le N \le 12$ Majorana fermions over a wide range of temperatures. Utilizing IBM's 127-qubit quantum processor, we perform benchmark computations for the dense SYK model with $N=6$, showing good agreement with exact results. The preparation of thermal states of a non-local random Hamiltonian with all-to-all coupling using the simulator and quantum hardware represents a significant step toward future computations of thermal out-of-time order correlators in quantum many-body systems.
\end{abstract}

\maketitle


\section{Introduction}

Certain quantum many-body systems exhibit properties that allow for a connection to problems involving black holes in a spacetime with one extra dimension. The theoretical framework that facilitates this mapping is the holographic duality, which in some cases is also called the AdS/CFT correspondence~\cite{Maldacena:1997re}. This duality relates a strongly interacting conformal field theory (CFT) involving a large number of degrees of freedom to a theory of weak gravity in an anti-de Sitter (AdS) background with one extra dimension. The Sachdev-Ye-Kitaev (SYK) model~\cite{Sachdev_1993, Kitaev:2015talk} is a quantum mechanical model in $(0+1)$-dimensions that consists of $N$ randomly interacting Majorana fermions. It exhibits an approximate conformal symmetry in the large $N$ and low-temperature limit $N \gg \beta \mathcal{J} \gg 1$, where $\beta$ is the inverse temperature while $\mathcal{J}$ is the disorder strength. Due to this relation, the SYK model has been studied extensively over the past decade~\cite{Maldacena:2016hyu, Kitaev:2017awl}. 

Due to its role in holography, a detailed study of both pure and thermal (or Gibbs) states of the SYK model is necessary. The ground (pure) state of the SYK model exhibits a volume law entanglement of the entanglement entropy with a known coefficient \cite{Huang2019}, which makes them classically hard to prepare~\cite{Passetti2023}. 
In addition, a defining property of the SYK model is that thermal four-point correlators, in particular out-of-time order correlators (OTOCs), saturate the exponent that measures the exponential growth of such correlators in the large $N$ and low-temperature limit~\cite{Maldacena:2015waa}. To measure these OTOCs, we have to prepare the relevant thermal state, which will be the focus of this work. While preparing low-energy states, which have been studied using large-scale classical simulations, our work aims to perform benchmark calculations using hybrid quantum/classical algorithms. The recent progress in quantum technologies and algorithms has led to various explorations of quantum computing in the context of fundamental physics~\cite{Chandrasekharan:1996ih, Banerjee:2012pg, Banuls:2019bmf, Shaw:2020udc, Bauer:2021gek, Liu:2022grf, Barata:2023jgd, Farrell:2024fit, Bauer:2023qgm, Briceno:2023xcm, ARahman:2022tkr, Meth:2023wzd, Grieninger:2024cdl, Araz:2022tbd, PhysRevA.109.062422, Araz:2022zxk, Araz:2022haf,Carrillo:2024chu,Khor:2023xar}. In elementary particle and nuclear physics, quantum computing promises to enable calculations that lie beyond the reach of classical simulations. In this work, we carry out hybrid classical/quantum simulations using variational quantum algorithms and small-scale benchmark simulations on quantum hardware. In recent years, several studies of the SYK model have been carried out using quantum computing methods~\cite{Garcia-Alvarez:2016wem} such as teleportation~\cite{Lykken:2024ypy}, computation of bosonic correlation functions~\cite{Luo:2017bno}, real-time dynamics~\cite{Asaduzzaman:2023wtd, Jha:2024vcw}, and the ground state preparation of the (coupled) SYK model~\cite{Su:2020zgc, Kim:2021ffs}. 

In earlier work, some of the authors focused on the real-time dynamics of the SYK model~\cite{Asaduzzaman:2023wtd} using quantum hardware. Next, we focus on the thermal state preparation in this work. The primary challenge here stems from the fact that thermal states are not pure. While quantum thermal state sampling algorithms have been developed in the fault-tolerant regime~\cite{Chen:2023cuc}, in this work, we focus on variational algorithms~\cite{Peruzzo2014, Bharti2022, Tilly:2021jem} that are promising in the near to intermediate term future. Alternative algorithms include purifying thermal states using ancillary qubits, canonical thermal pure quantum states, and coupling the system to a thermal bath~\cite{Sugiura:2013pla, Powers:2021nqh, Bravyi:2021pwq}. See also Refs.~\cite{Cohen:2021imf, Zhou:2021kdl, Mueller:2021gxd, Czajka:2021yll, Honda:2021aum,  Barata:2022wim, Davoudi:2022uzo, Xie:2022jgj, Martin:2023gbo, Florio:2023dke, Lee:2023urk, Saroni:2023uob, Gustafson:2023xpe, PhysRevA.109.062422, Florio:2024aix, Qian:2024xnr, Ebner:2024mee, Davoudi:2024osg, Olsacher:2024wil,Cirigliano:2024pnm} for recent work on different aspects of thermal states in the context of fundamental physics. 

In recent years, methods such as Variational Quantum Eigensolver (VQE)~\cite{Peruzzo2014} and the Quantum Approximate Optimization algorithm (QAOA)~\cite{Farhi:2014ych} have been used for the ground state preparation of gapped quantum systems. These hybrid quantum/classical algorithms comprise a parametrized quantum circuit as an ansatz of the ground state and a classical optimization loop that minimizes a cost function. For ground state preparation, the cost function is the expectation value of the Hamiltonian, whereas for thermal states, the Helmholtz free energy is minimized. In this paper, we employ the algorithm presented in Ref.~\cite{Selisko:2022wlc}. Aside from studying the thermal states of the (dense) SYK model, we also consider its sparsified version~\cite{Garcia-Garcia:2020cdo, Xu:2020shn, Garcia-Garcia:2023jlu, Orman:2024mpw}. Several random Hamiltonians like the SYK model have also been investigated from the complexity theory point of view~\cite{Hastings:2021ygw, Chen:2023lhm}. While the low-energy state preparation of general local Hamiltonians is Quantum Merlin Arthur (QMA) hard in the worst case, it was demonstrated in Ref.~\cite{Chen:2023lhm} that efficient complexity guarantees exist for the average case of certain random Hamiltonians. In addition, random Hamiltonians have played an essential role in quantum chaos and in nuclear physics, where they serve as models of heavy nuclei~\cite{Mitchell:2010um} going back to the observation of Wigner (famous surmise) that the spectral gap distribution in nuclei of heavy atoms can be described by random matrix ensembles. 

This paper is organized as follows. In Sec.~\ref{sec:SYK}, we introduce the full and sparse SYK model. In Sec.~\ref{sec:qVQT}, we review the variational thermal state preparation algorithm used in this work. In Sec.~\ref{sec:Results}, we present results from the classical simulator and IBM's quantum hardware. We conclude and present an outlook in Sec.~\ref{sec:Conclusion}.

\begin{figure*}
    \centering
    \includegraphics[width=0.7\linewidth]{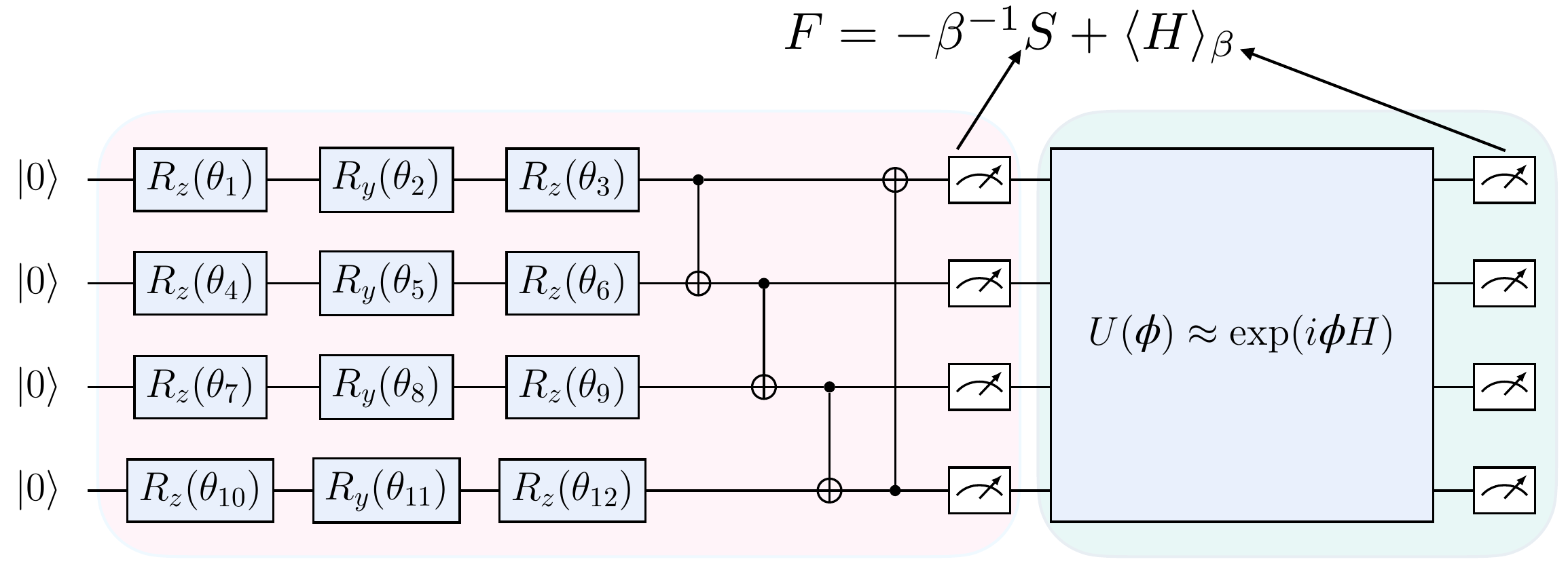}
    \caption{\it Illustration of the variational quantum thermal state preparation algorithm used in this work, adapted from Ref.~\cite{Selisko:2022wlc} for the $N=8$ version of the model. The left circuit represents one layer (with $3n$ parameters) used to variationally determine the classical Shannon entropy $S$ through intermediate measurements. The resulting measurements in the computational basis are then taken as the initial state of the second variational circuit that determines the energy of the Hamiltonian $\braket{H}_\beta$. Both the entropy $S$ and the energy $\braket{H}_\beta$ are used to calculate the loss function, which is given by the Helmholtz free energy $F$.\label{fig:circuit_algo}}
\end{figure*}

\section{The dense and sparse SYK model~\label{sec:SYK}}

SYK models are a class of fermionic models with a disorder average. These types of many-body systems have been studied for several decades due to their role in the physics of spin glasses and non-Fermi liquids in strongly correlated systems. One example is the Sachdev-Ye model~\cite{Sachdev:1992fk}, studied in the context of quantum Heisenberg magnets with random Gaussian and infinite-range interactions. These models were often revisited in the light of the holographic conjecture~\cite{Sachdev:2010uj}. However, it was challenging to find a concrete model exhibiting holographic features. This changed in 2015 when Kitaev showed that the four-body quenched disorder model has holographic properties when quartic Majorana operators replace the spin operators in the limit of a large number of fermions. In particular, it was shown to exhibit an approximate conformal symmetry (time reparametrization) in the low-temperature, large-$N$ limit where the holographic description can emerge. 

This model is now known as the SYK model. In this section, we describe this model and its sparsified version. In subsequent sections, we will study the thermal states of both models using a variational thermal quantum algorithm. The Hamiltonian for the SYK model with $N$-Majorana fermions is given by
\begin{equation}\label{eq:SYK_main} 
H = \frac{1}{4!} \sum_{ijkl}^{N} J_{ijkl}~\chi_{i} \chi_{j} \chi_{k} \chi_{l}\,.
\end{equation}
Here, $\chi_i$ denote Majorana fermions $\chi_i^\dagger=\chi_i$ that satisfy the anti-commutation relation $\{\chi_i, \chi_j\} =  \delta_{ij}$. The sum runs over all possible quartic interactions. The random couplings $J_{ijkl}$ are drawn from a Gaussian distribution with mean and variance given by
\begin{equation}\label{eq:Jijkl}
\overline{J_{ijkl}}=0\,,\quad \overline{J_{ijkl}^2}=\frac{3! \mathcal{J}^2}{N^3} \,.
\end{equation}
This model can also be considered with interaction terms involving more than four fermions. Such versions are known as the $q$-SYK model. However, here, we will restrict ourselves to the version given in Eq.~\eqref{eq:SYK_main} with $q=4$. We also set the disorder strength to $\mathcal{J} = 1$, see Eq.~(\ref{eq:Jijkl}). For this work, we choose dimensionless parameters and refer to $\beta \mathcal{J}$ simply as $\beta$. 

One of the challenges of the full or dense SYK model is that the number of terms in the Hamiltonian grows as $\sim N^4$. To reduce the computational cost of classical or quantum simulations of the SYK model, a sparsified version was introduced in Ref.~\cite{Garcia-Garcia:2020cdo, Xu:2020shn}. See also Refs.~\cite{Su:2020zgc, Garcia-Garcia:2020cdo, Jha:2024vcw}. \CHANGE{In the large-$N$ limit, the SYK model has ${{N}\choose{4}}  \sim N^{4}/24$ terms. The number of terms in the Hamiltonian per Majorana fermion is $k=N^{3}/24$, which is also called the average connectivity of the corresponding hypergraph. When $k \ll N^{3}/24$, the resulting Hamiltonian is referred to as the sparse SYK model. The value of $k$ can be chosen as small as possible to reduce the computational cost but sufficiently large to retain a holographic interpretation, see Ref.~\cite{Orman:2024mpw}. 
In this work, we consider the sparse SYK model with a fixed sparsity of $k = 8.7$.} 
Following Ref.~\cite{Xu:2020shn}, the sparsified SYK model is obtained by modifying the Hamiltonian in Eq.~\eqref{eq:SYK_main} as
\begin{equation}
H = \frac{1}{4!} \sum_{ijkl}^{N} p_{ijkl} J_{ijkl}~\chi_{i} \chi_{j} \chi_{k} \chi_{l} \,.
\label{eq:SYK_main_sparse} 
\end{equation}
Here, the additional factor $p_{ijkl}$ is either 0 or 1, depending on whether we remove or keep the respective term in the Hamiltonian. The terms of the original SYK model are removed with probability $1-p$ and retained with probability $p$. The random couplings $J_{ijkl}$ are now taken from a modified Gaussian distribution with mean and variance given by
\begin{equation}
    \overline{J_{ijkl}}=0\,,\quad \overline{J_{ijkl}^2}=\frac{3! \mathcal{J}^2}{pN^3} \,.
\end{equation}
The variance of the couplings $J_{ijkl}$ is more significant compared to Eq.~\eqref{eq:Jijkl}, which ensures that the energy scales are comparable to the original SYK model. The probability with which the terms are retained is related to the average connectivity between fermions $k$ as $p \approx 24k/N^{3}$. Therefore, the variance of the sparse SYK model is $\approx 3! \mathcal{J}^2/24k$ and independent of the number of fermions $N$ for a given value of $k$. The dense or full SYK model corresponds to $p=1$. It was shown in Ref.~\cite{Garcia-Garcia:2023jlu} that until one goes below the percolation limit, the Lyapunov exponent does not depend strongly on the sparsity of the model. 

The product of Majorana fermions in Eqs.~\eqref{eq:SYK_main} and~\eqref{eq:SYK_main_sparse} can be mapped to Pauli strings using a Jordan-Wigner transformation~\cite{Jordan:1928wi}. Here, $N$ Majorana fermions can be represented by $\lceil{N/2}\rceil$ qubits. We will use $n = N/2$ for the total number of qubits from here on since we only consider an even number of Majorana fermions. We refer the reader to Refs.~\cite{Garcia-Alvarez:2016wem, Asaduzzaman:2023wtd} for more details. 

For both the full and sparse SYK model, we aim to construct an approximate thermal state $\rho$ at an inverse temperature $\beta$ that is described by the thermal density matrix
\begin{equation}
\rho_{\beta} =  \frac{e^{-\beta H}}{Z}\,.
\end{equation} 
Here $Z = \sum_{i} e^{-\beta E_{i}}$ is the canonical partition function and the Hamiltonian $H$ is either given in Eq.~\eqref{eq:SYK_main} or \eqref{eq:SYK_main_sparse}. We will employ variational algorithms as described in the next section to access properties of the thermal state. We quantify the performance of the variational state preparation by comparing the resulting density matrix $\rho$ to the thermal state $\rho_{\beta}$ obtained using exact diagonalization. To assess how well we can approximate the target density matrix, we calculate the Uhlmann-Jozsa fidelity~\cite{Jozsa1994}
\begin{equation}
{\cal F}(\rho_{\beta},\rho) = {\mbox{Tr}} \Big(\sqrt{\rho_{\beta}^{1/2} \rho~ \rho_{\beta}^{1/2}}\Big) \,.
\label{eq:fidelity} 
\end{equation}
The fidelity $0\leq {\cal F}\leq 1$ is symmetric by definition and measures the degree of distinguishability between the two density matrices. The fidelity is equal to 1 if and only if the density matrices are the same.

The degeneracy of the ground state of the SYK model depends on the number of fermions $N$. When $N\, \text{mod 8} = 0$, the ground state is unique. Otherwise, it is doubly degenerate. For the SYK model, the energy spectrum is dense with a very small gap between the ground and excited states~\cite{Maldacena:2016hyu}. In addition, the ground state entanglement entropy follows a volume law. These aspects make the variational state preparation of this model challenging, and a significant number of trainable parameters are needed, as already observed in Refs.~\cite{Kim:2020luc, Kim:2021ffs}. Our results constitute the first attempt at variational quantum computations of the thermal states of the SYK model. In the next section, we will describe the algorithm used in this work and then discuss the results obtained from the simulator and those using IBM's quantum platform for $N=6$ for one instance of the dense SYK model.

\section{Variational thermal quantum algorithm~\label{sec:qVQT}}

Thermal states are mixed states that are described by a density matrix. They are generally more challenging to study compared to ground states at zero temperature. On the other hand, at very high temperatures, the thermal state is a maximally mixed state described by the density matrix $\rho \propto \mathds{1}$. At high temperatures, thermal states are thus well approximated by a classical ensemble. The extreme limits of $\beta\to 0$ and $\beta\to\infty$ are computationally more straightforward to prepare compared to finite inverse temperature $\beta$. The finite-$\beta$ case is the relevant regime for pertinent various studies of the SYK model and other quantum many-body systems.

The loss function of variational quantum algorithms for the ground state preparation is the expectation value of the Hamiltonian $\braket{H}$, which needs to be minimized. To prepare a thermal state at finite $\beta$, which is described by the density matrix $\rho_\beta$, we minimize instead the Helmholtz free energy $F$, which is given by
\begin{equation}\label{eq:FreeEnergy}
    F = \langle H \rangle_{\beta} - T\cal{S}\,.
\end{equation}
Here, $T=1/\beta$ is the temperature, $\cal{S}$ is the von Neumann entropy given by $\cal{S} = - \text{Tr}[\rho_\beta \log \rho_\beta]$ and $\braket{H}_\beta=\text{Tr}[\rho_\beta H]$ is the expectation value of the Hamiltonian.

\begin{figure*}
    \centering
    \subfigure[$N=6$, 10 instances]{\includegraphics[width=0.39\linewidth]{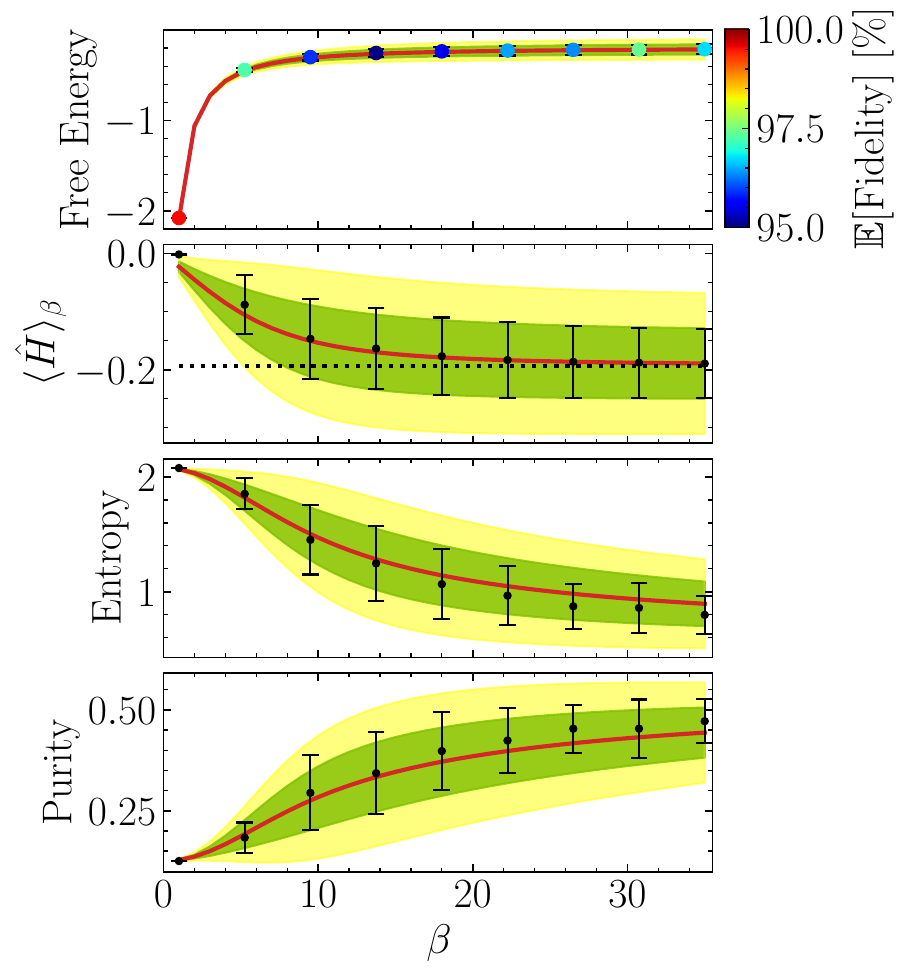}\label{fig:res-dense-n6}}\quad\quad\quad\quad
    \subfigure[$N=8$, 10 instances]{\includegraphics[width=0.39\linewidth]{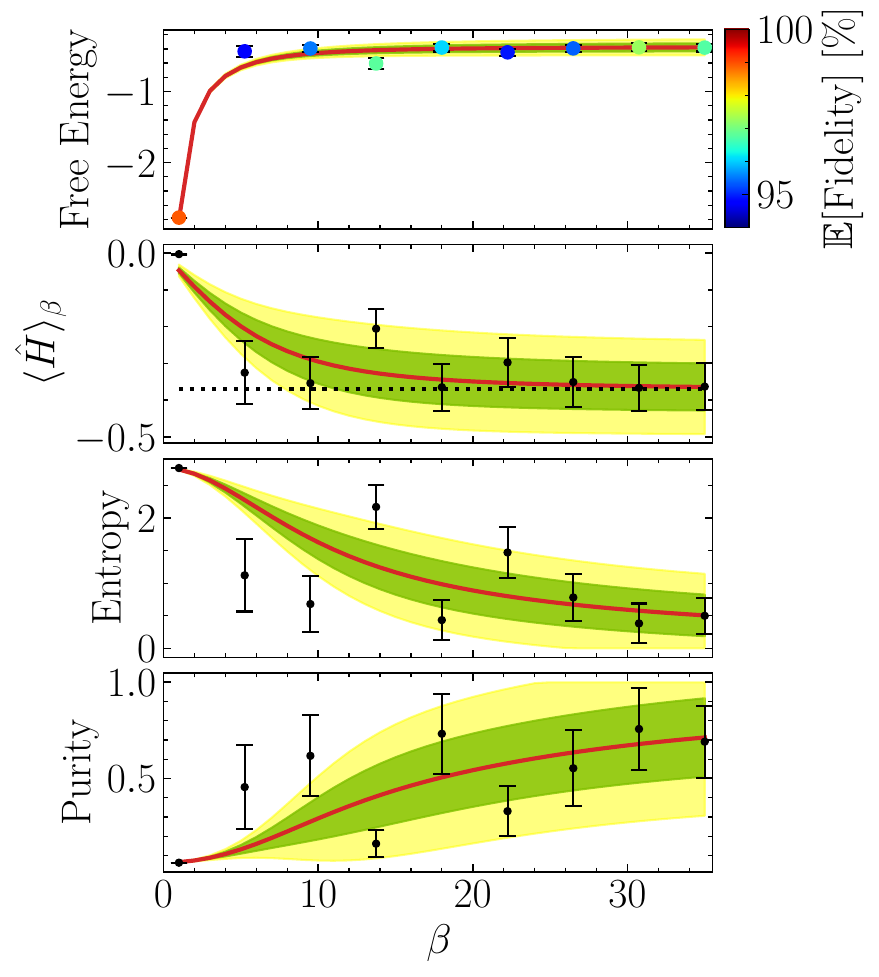}\label{fig:res-dense-n8}}
    \subfigure[$N=10$, 10 instances]{\includegraphics[width=0.39\linewidth]{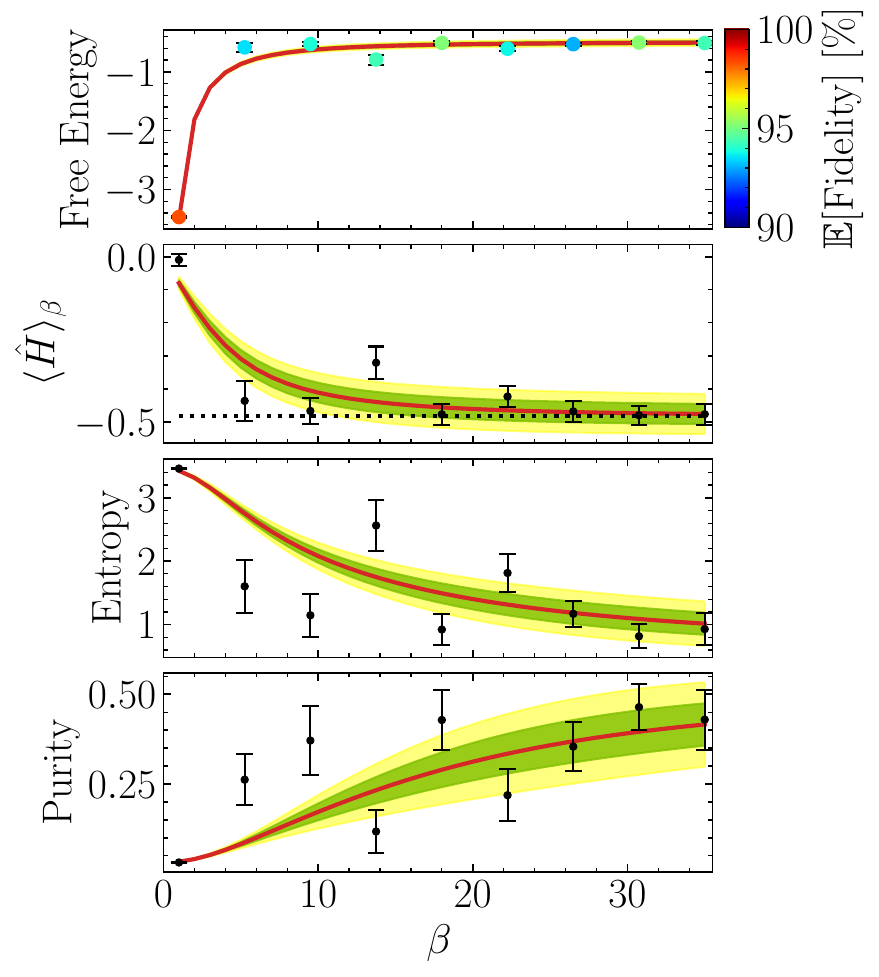}\label{fig:res-dense-n10}}\quad\quad\quad\quad
    \subfigure[$N=12$, 5 instances]{\includegraphics[width=0.39\linewidth]{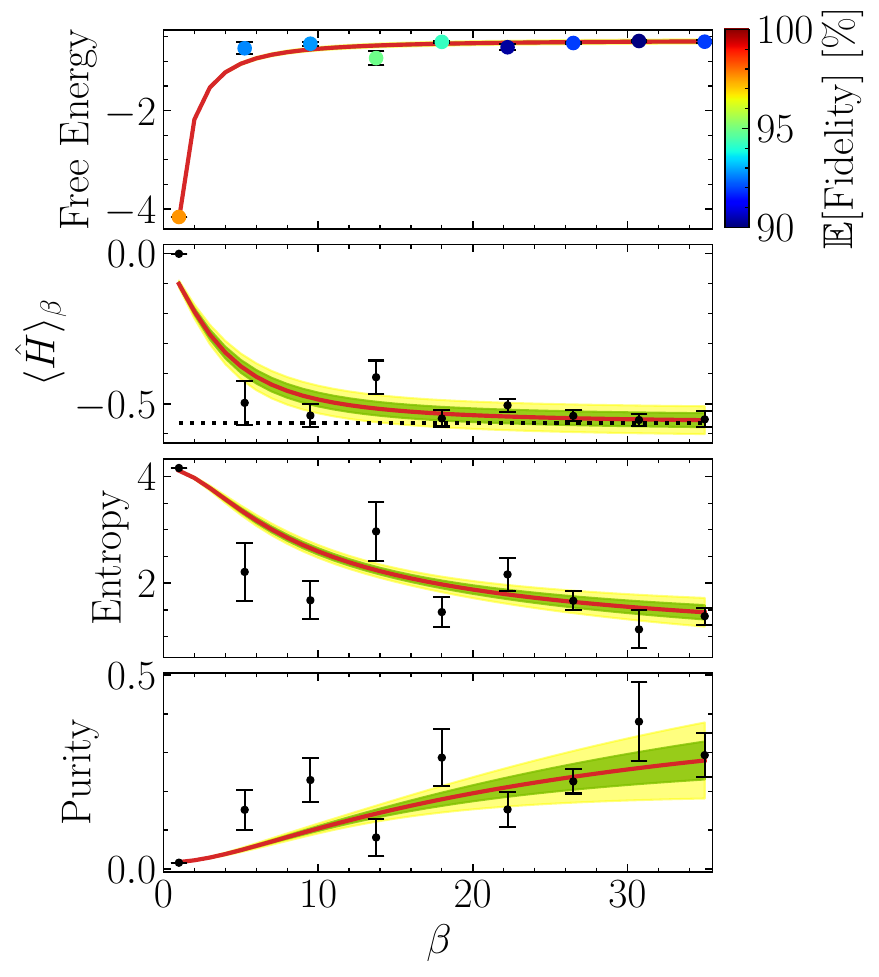}\label{fig:res-dense-n12}}
    \caption{\it The results for the dense SYK model for $6 \leq N \leq 12$. The red line shows the average result of observables (computed from the exact thermal density matrix) over the instances, and the shaded regions denote the $\pm 1\sigma,\ \pm 2\sigma$ deviations for the disorder average over multiple instances of the Hamiltonian. The black dots represent results from the variational algorithm where error bars show one standard deviation over all instances. The dashed black lines denote the ground state energy.}
    \label{fig:res_dense}
\end{figure*}

\begin{figure*}
    \centering
    \subfigure[$N=6$, 10 instances]{\includegraphics[width=0.39\linewidth]{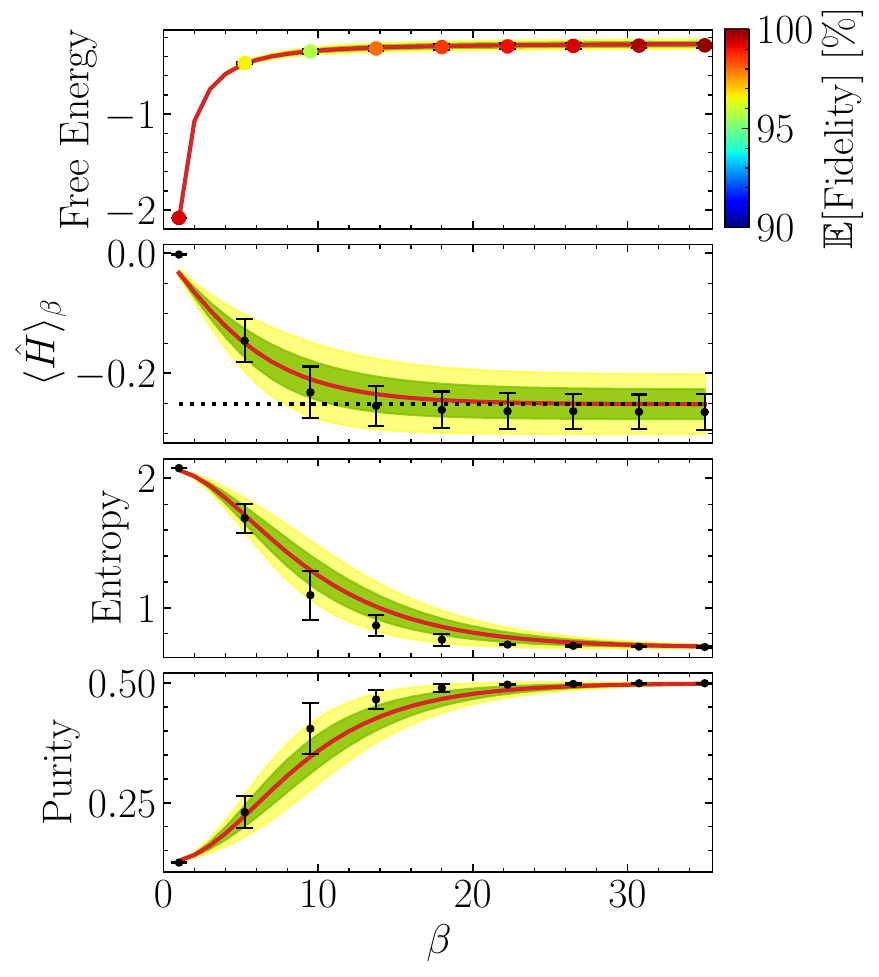}\label{fig:res-sparse-n6}}\quad\quad\quad\quad
    \subfigure[$N=8$, 10 instances]{\includegraphics[width=0.39\linewidth]{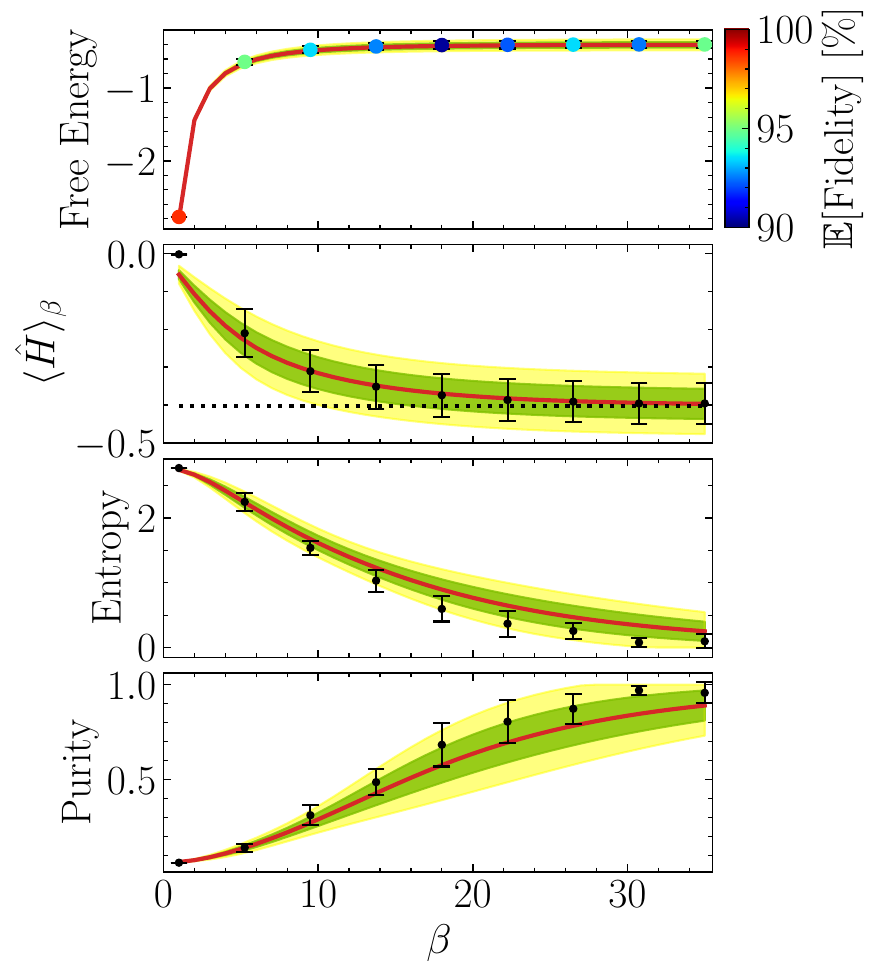}\label{fig:res-sparse-n8}}
    \subfigure[$N=10$, 10 instances]{\includegraphics[width=0.39\linewidth]{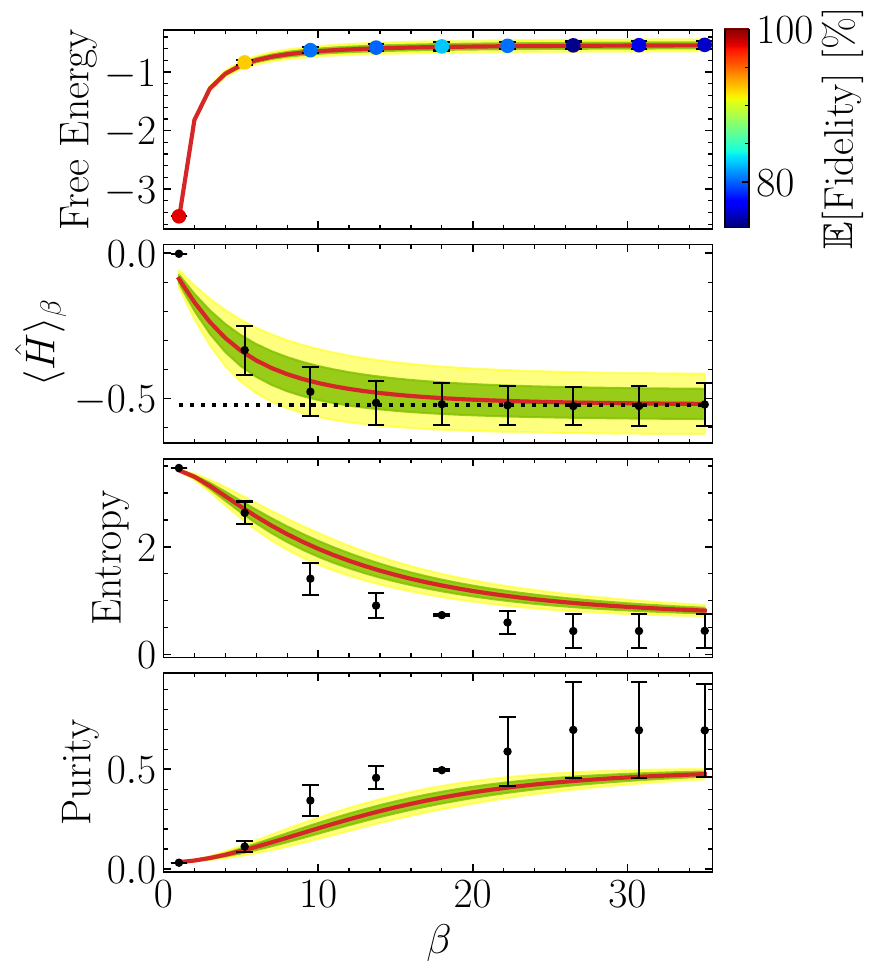}\label{fig:res-sparse-n10}}\quad\quad\quad\quad
    \subfigure[$N=12$, 5 instances]{\includegraphics[width=0.39\linewidth]{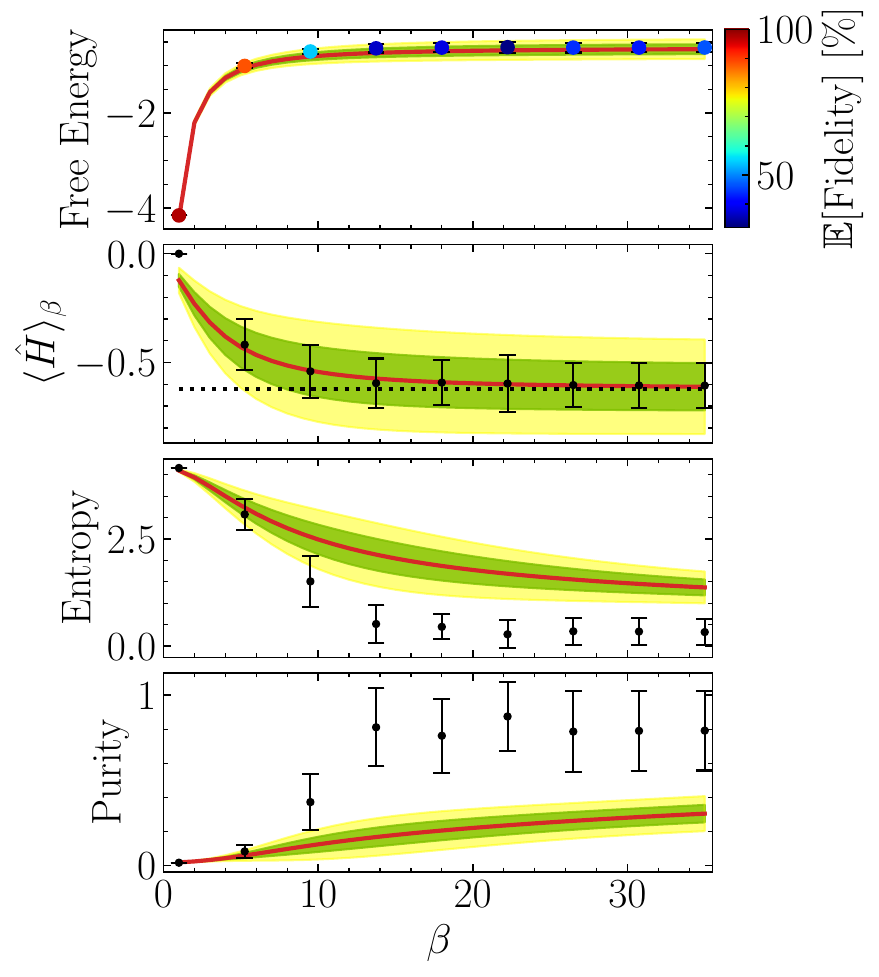}\label{fig:res-sparse-n12}}
    \caption{\it The results for the sparse SYK model with $k=8.7$ for $6 \leq N \leq 12$. The red line shows the average result of observables (computed from the exact thermal density matrix) over the instances, and the shaded regions denote the $\pm 1\sigma,\ \pm 2\sigma$ deviations for the disorder average over multiple instances of the Hamiltonian. The black dots represent results from the variational algorithm where error bars show one standard deviation over all instances. The dashed black lines denote the ground state energy.}
    \label{fig:res_sparse}
\end{figure*}

In this work, we adopt the variational thermal state preparation algorithm introduced in Ref.~\cite{Selisko:2022wlc}. It consists of a sequence of two variational quantum circuits as illustrated in Fig.~\ref{fig:circuit_algo} for four representative qubits. We initialize all qubits in the $\vert 0 \rangle^{\otimes n}$ state.  In contrast to the ground state preparation, thermal states cannot be constructed with purely unitary circuits since they are mixed states, i.e. a probabilistic mixture of different pure states. This is reflected by the intermediate measurement of the qubits as illustrated in Fig.~\ref{fig:circuit_algo}. Before the intermediate measurement, the left circuit represents one layer of the variational circuit that we refer to as VQC$_1$. Each layer consists of $n$ two-qubit gates and $3n$ rotation gates, representing a hardware-efficient ansatz~\cite{Peruzzo2014}. For the single-qubit gates, we choose a sequence of rotation gates $R_z\, R_y\, R_z$ that depends on variational parameters. The single-qubit gate sequence is followed by CNOT entangling gates. We denote the set of variational parameters of VQC$_1$ by ${\bm \theta}$, the unitary operation where the gates in all layers of the variational circuit are applied is denoted by $U_1({\bm \theta})$, and the resulting state is given by $\ket{\psi({\bm \theta})}=U_1({\bm\theta})\ket{0}^{\otimes n}$. For the numerical calculations presented in the next section, we increase the number of layers until we reach a fidelity of ${\cal F}>0.9$, see Eq.~\eqref{eq:fidelity}. After VQC$_1$ is executed, we measure the qubits in the computational basis to determine the probabilities associated with the $i^{\rm th}$ orthonormal basis vector. The targeted thermal state can be written in terms of the eigenstates $\vert \varphi_i \rangle$  of the Hamiltonian $H$ and the corresponding energies $E_i$ as
\begin{equation}\label{eq:rhoH}
 \rho_\beta = \sum_{i} p_i |\varphi_{i} \rangle \langle \varphi_{i}|\,,
\end{equation}
where the sum runs over $2^{n}$ states. The probabilities are given by $p_i =  e^{-\beta E_i}/Z$ where $Z = \sum_i e^{-\beta E_i}$ is the canonical partition function. The trace of the density matrix is normalized to 1. After applying VQC$_1$, the intermediate measurement uses the computational basis. We denote the basis vectors by $\ket{b_i}$, and we can write the density matrix before the intermediate measurement as
\begin{align}
\rho_{\rm{VQC}_1} &= \vert \psi({\bm \theta}) \rangle \langle \psi({\bm \theta}) \vert \nonumber \\ 
 & = \Big( \sum_{i} a_{i}({\bm \theta}) \vert b_{i} \rangle  \Big) \Big( \sum_{j} a^{*}_{j}({\bm \theta}) \langle b_{j} \vert \Big)\,. 
\end{align} 
The intermediate measurement reduces the density matrix to its diagonal components on a computational basis, and we obtain a probability distribution. The density matrix after the measurement can, therefore, be written as
\begin{equation}
    \tilde\rho_{{\rm VQC}_1} = \sum_i\vert a_{i}({\bm \theta}) \vert^{2} \ket{b_i}\bra{b_i}\,.
\end{equation}
Using the measurement outcome, we can determine the classical Shannon entropy
\begin{equation}\label{eq:ShannonEntropy}
    S = - \sum_{i=1}^{2^{n}} p_{i}({\bm \theta}) \log p_{i}({\bm \theta})\,.
\end{equation}
Different than in Eq.~\eqref{eq:rhoH}, $p_{i}({\bm \theta}) = \vert a_{i}({\bm \theta}) \vert^{2}$ corresponds to the probability of the $i^{\rm th}$ basis state in the ensemble. To calculate the free energy in Eq.~\eqref{eq:FreeEnergy}, we approximate the von Neumann ${\cal S}$ with the Shannon entropy $S$. Note that the von Neumann entropy would require access to the entire density matrix, whereas the Shannon entropy can be calculated from its diagonal entries. The Shannon entropy is equal to von Neumann entropy only in the Schmidt basis. \CHANGE{This simplification can sometimes result in a less accurate construction of the thermal state.
However, even for quantum computers an explicit computation of von Neumann entropy is possibly not efficient. An optimal thermal state algorithm potentially has to converge to a convex mixture, which is the closest approximation to the Schmidt basis where the Shannon and von Neumann entropy agree}. The measurement outcomes in the computational basis of VQC$_1$ are now taken as the initial state for the second part of the algorithm. We denote the unitary operation of the set $\left\{\text{VQC}_2\right\}$ by $U_2({\bm \phi})$, where ${\bm \phi}$ denotes the  variational parameters. Different than for VQC$_1$, we start with a Hamiltonian-based ansatz~\cite{Gard_2020}, i.e. the variational circuit consists of gates that appear in the Trotterized time evolution operator associated with the Hamiltonian given in Eq.~\eqref{eq:SYK_main}.
\begin{figure*}
    \centering
    \includegraphics[width=0.9\linewidth]{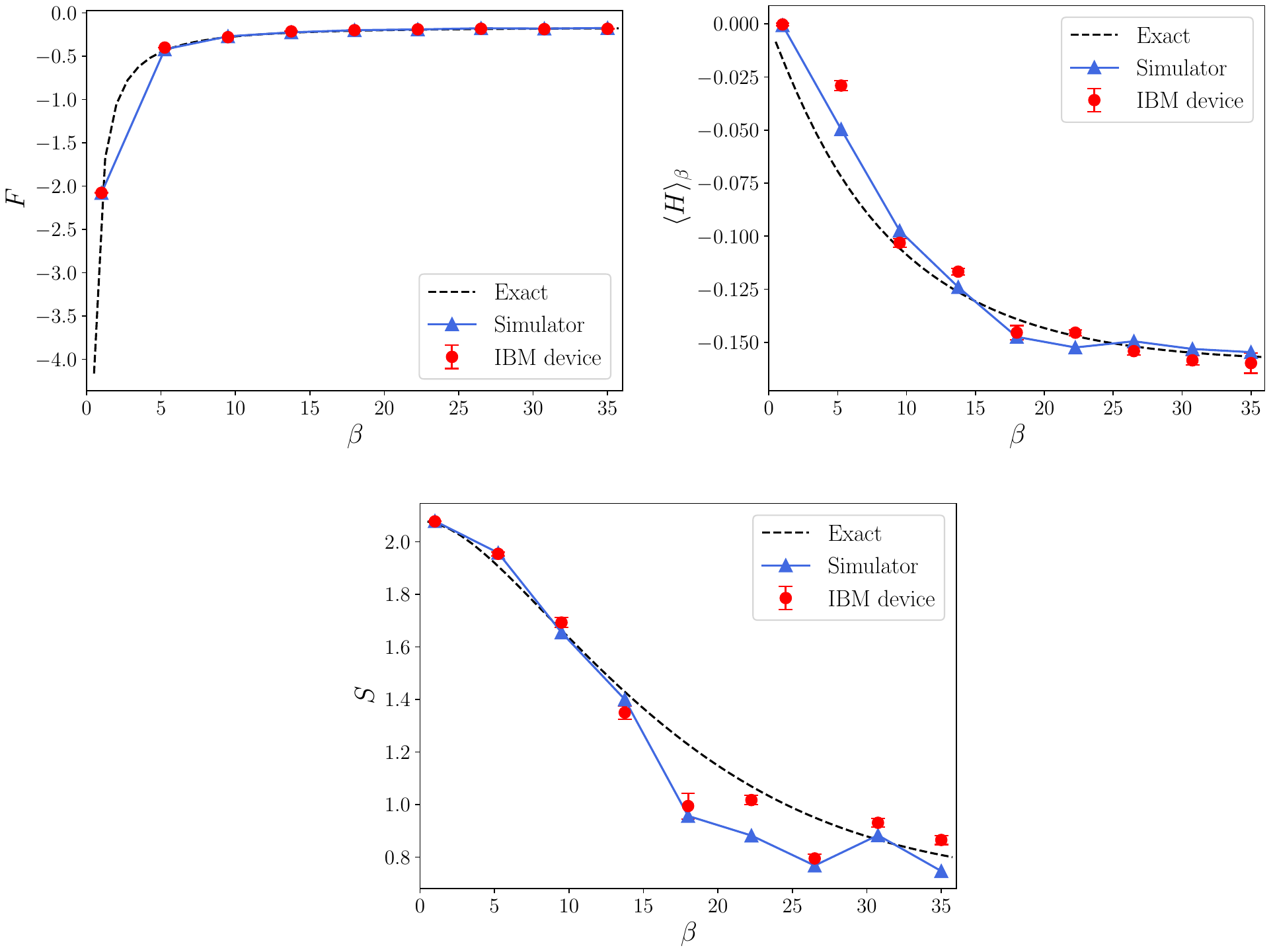}
    \caption{\it Numerical results for the thermal state preparation of a single instance of the $N=6$ dense SYK model. The three panels show the free energy $F$, the energy $\langle H\rangle_{\beta}$, and the entropy $S$ as a function of the inverse temperature $\beta$. We show the exact results and the results from the classical and quantum simulations using IBM's quantum hardware.\label{fig:N6hardware}}
\end{figure*}
However, we observed that a purely Hamiltonian-based ansatz is not efficient enough to achieve the desired fidelity. Therefore, we cluster different terms of the Hamiltonian to reduce the overall circuit depth. We employ the techniques described in Ref.~\cite{Murairi:2022zdg} to find the diagonalizing circuit for each cluster where all terms commute. This approach significantly improved the convergence of our algorithm. The second part of the thermal state preparation algorithm implements the unitary operation that maps the basis states to a superposition of states, maintaining orthonormality. This gives the following density matrix
\begin{equation}
\rho_{\rm{VQC}_2} = \sum_{i} |a_{i}({\bm \theta})|^{2} \vert \Psi_{i}({\bm \phi}) \rangle \langle 
\Psi_{i}({\bm \phi}) \vert\,,
\end{equation}
which approximates the thermal density matrix $\rho_\beta$. Here $\ket{\Psi_i({\bm \phi})}=U_2({\bm \phi})\ket{b_i}$ 
denotes the state obtained by applying $\left\{\text{VQC}_2\right\}$ after the intermediate measurement. This allows us to compute $\langle H \rangle_{\beta} = \mathrm{Tr}[\rho_{{\rm{VQC}_2}} H]$, and thus we obtain the second term needed to determine the free energy $F$ in Eq.~\eqref{eq:FreeEnergy}. We can approximate the thermal state by minimizing $F$ using a classical optimizer to determine the parameters ${\bm\theta},{\bm\phi}$. To assess the performance of the algorithm in preparing the thermal state, we measure the fidelity ${\cal F}(\rho_\beta,\rho_{\rm VQC_2})$ defined in Eq.~\eqref{eq:fidelity}. In addition to the free energy, entropy, and the energy expectation value, we also measure the purity of the thermal state defined as:
\begin{equation}
    P = {\rm Tr}[ \rho_{{\rm VQC}_2}^2]\,.
\end{equation}
Based on whether the state is pure or maximally mixed, the purity $P$ is bounded by $1/d \le P \le 1$ where $d=2^n$ is the dimension of the Hilbert space. 

\section{Numerical results~\label{sec:Results}}

In this section, we present numerical results using both a classical simulator and IBM's quantum hardware. Using the classical simulator, we explore a wide range of $N$ and multiple instances for both the dense and sparse SYK model. However, when using the quantum hardware, we limit ourselves to a single instance of the $N=6$ dense SYK model.

\subsection{Simulator and optimized parameters}

We employ the variational quantum thermal state algorithm introduced in Ref.~\cite{Selisko:2022wlc}. For both the dense and sparse SYK model, we consider two circuits using strongly entangling layers~\cite{Schuld:2018ahn} for reconstructing the entropy and a Hamiltonian-based ansatz constructed using a first-order Trotter approximation, as described also in the previous section. The single-qubit gates contain the variational parameters to reconstruct the ground state energy. A schematic illustration of our approach is shown in Fig.~\ref{fig:circuit_algo}. For the numerical simulations presented here, we use \textsc{PennyLane} (version 0.35.1)~\cite{bergholm2020pennylane} implementation with the \textsc{Jax} (version 0.4.23)~\cite{jax2018github} extension to utilize gradients associated with the parametrized gates for optimization. The optimization is performed using \textsc{SciPy}'s (version 1.10.0)~\cite{2020SciPy-NMeth} Sequential Least Squares Programming (SLSQP) method. More details of the variational algorithm are described in Appendix~\ref{app:var}.

The required fidelity of 90\% is achieved by incrementally increasing the layer count of the circuits with a maximum of ten layers for either circuit. In Fig.~\ref{fig:res_dense}, we show the results for the dense SYK model for different numbers of fermions $N$. Each simulation was carried out for 10 instances, except for $N=12$, where we were limited to 5 instances due to increased computational complexity. In each panel of Fig.~\ref{fig:res_dense}, the results for the free energy, the expectation value of the Hamiltonian, the entropy, and the purity distributions are shown using exact diagonalization and the variational algorithm. The red line in each plot shows the mean of the exact results, and the green (yellow) region shows the $1\sigma$ ($2\sigma$) deviation from the mean. The black dots in each panel show the variationally reconstructed results. The error bars represent one standard deviation for the given number of instances of the model. The black dots in each free energy plot have been replaced with a color map where the color represents the achieved mean fidelity. For $N=6$, we observe good agreement between the exact results and those obtained with the variational algorithm. All error bars remain in the green region, representing a one-sigma deviation from the exact mean results. However, as we increase the number of fermions $N$, we observe significant fluctuations in the intermediate-$\beta$ region. 

In Fig.~\ref{fig:res_sparse}, we show the same results for the sparse SYK Hamiltonian. Although we observe an accurate reconstruction of the energy distribution for each number of fermions, the entropy distribution is consistently lower than the exact results. The deviation grows as the number of fermions is increased. On the other hand, the purity distribution is above the exact results, with the deviation again increasing with the number of fermions. We observe that small temperatures agree well with the ground state energy beyond $\beta \gtrsim 20$. In Appendix~\ref{app:var}, we describe in more detail the resource requirements of the dense and sparse SYK model. For comparison, we additionally describe the construction of the thermal state from the purified Gibbs state or Thermofield Double (TFD) state in Appendix~\ref{app:TFD}.

\subsection{Quantum hardware results}

We now turn to the discussion of the quantum hardware implementation for $N=6$. After finding the set of parameters for the variational circuits $\text{VQC}_1({\bm \theta})$, $\left\{\text{VQC}_2({\bm \phi})\right\}$ that minimize the free energy, we can measure the properties of that state using quantum hardware for small values of $N$. Given current hardware limitations, we choose $N=6$ and measure the Shannon entropy $S$, the expectation value of the Hamiltonian $\langle H\rangle_{\beta}$, and the free energy $F$ defined in Eq.~\eqref{eq:FreeEnergy} for different values of $\beta$. For $N=8$, we could not obtain accurate results from the quantum hardware due to the higher resource requirements. The measurement of the Shannon entropy using the quantum hardware is relatively straightforward as it only requires the diagonal entries of the density matrix after applying $\text{VQC}_1$ as described in Eq.~\eqref{eq:ShannonEntropy}. 

The calculation of the expectation value of the Hamiltonian $\langle H \rangle_{\beta}$ for a fixed value of $\beta$, on the other hand, requires the off-diagonal elements of the density matrix after the application of $\left\{\text{VQC}_2\right\}$. Moreover, as described in Sec.~\ref{sec:qVQT}, to measure the expectation value of the Hamiltonian, we need $2^{n}$ circuits -- one for each computational basis state. This scaling, combined with the large number of measurements required to carry out the tomography, makes the measurement of $\langle H\rangle_{\beta}$ computationally challenging. A possibility to reduce the computational cost is using recently developed techniques such as classical shadow tomography~\cite{Huang:2020tih}. One could also reduce the number of $\left\{\text{VQC}_2\right\}$ circuits by analyzing the probability distribution of the intermediate measurement and only constructing the full circuit for basis states that contribute significantly to the Shannon entropy. It would be interesting to study how much these techniques can reduce the number of $\left\{\text{VQC}_2\right\}$ circuits for the SYK model as a function of $N$. For the cases considered here, i.e. $N < 14$, we do not require classical tomography of the output density matrix since it is cheap enough to transform the Hamiltonian to a diagonal (computational) basis and measure $\braket{H}_{\beta}$. However, going to larger values of $N$, it would be more efficient to consider  
efficient tomography techniques, and we leave that for future work.

We use IBM's \texttt{eagle\textunderscore r3} processors for preparing the states and measuring the relevant observables. The native two-qubit gate of these processors is the Echoed Cross-Resonance (\texttt{ECR}) gate defined as $1/\sqrt{2}(\mathbb{I}\otimes X-X\otimes Y)$. Due to the qubit connectivity (heavy-hex) topology of the quantum processor, the first and the third qubit (for $N=6$) are not connected, and we employ extra two-qubit \texttt{SWAP} gates. The two-qubit circuit depths of both the $\text{VQC}_1$ and $\left\{\text{VQC}_2\right\}$ for the $N=6$ dense SYK model are shallow enough to not require error mitigation protocols in addition to those that are available with IBM software. We obtain short circuit depths by compiling using the software package introduced in Ref.~\cite{osti_1785933}. For the built-in error mitigation schemes, we use the M3 protocol. We account for readout errors and employ zero noise extrapolation (ZNE). Here, any single two-qubit gate is replaced with three gates, and we extrapolate to the zero-noise limit for the measured observables. The results for a single instance of the $N=6$ dense SYK model are shown in Fig.~\ref{fig:N6hardware}. The error bars are obtained from five separate runs for each $\beta$ value with 8192 shots. While some observables show more significant deviations than others, overall, we find good agreement between the quantum simulations and the classical results.

\section{Conclusions and outlook~\label{sec:Conclusion}}

In this work, we investigated the use of variational quantum algorithms to prepare the thermal state of the dense and sparse SYK models. We used classical simulations to consider a wide range of temperatures for up to $N=12$ Majorana fermions. The variational algorithm minimizes the free energy and consists of two variational circuits that are linked by an intermediate measurement. We found a good agreement between the free energy, the Hamiltonian expectation value, and the fidelity of the prepared state by comparing them with the result using exact diagonalization. In addition to the results from the quantum simulator, we also prepared the thermal state for one instance of the $N=6$ dense SYK model for various temperatures using IBM's quantum hardware. After applying suitable error mitigation techniques, we achieved good agreement with the classical simulations. A significant challenge of the approach considered in this work is its scalability. Due to the lack of computational resources and the increased complexity for a large number of fermions, we were not able to extend our studies to $N \ge 14$. To extend quantum algorithms to larger values of $N$, the use of efficient state tomography algorithms such as classical shadows appears to be essential. It is likely that improving the algorithm, particularly the computation of the entropy using the first variational circuit, could improve the overall accuracy, which will be addressed in future work. With better scalability and accuracy. It would be compelling to compute thermal correlation functions, most notably the four-point out-of-time-order correlators~\cite{Green:2021fcl}, to further investigate quantum chaos in this model of quantum gravity. These explorations will be pursued in future work.

\section*{Acknowledgements}
R.G.J is supported by the U.S. Department of 
Energy, Office of Science, Office of Nuclear Physics with contract number DE-AC05-06OR2317 and 
U.S. Department of Energy, Office of Science, 
National Quantum Information Science Research Centers, 
Co-design Center for Quantum Advantage under contract number DE-SC0012704. JYA and FR are supported by the U.S. Department of Energy, Office of Science, Contract No.~DE-AC05-06OR23177, under which Jefferson Science Associates, LLC operates Jefferson Lab and in part by the DOE, Office of Science, Office of Nuclear Physics, Early Career Program under contract No. DE-SC0024358. BS is partly supported by the U.S. Department of Energy, Office of Science, Office of High Energy Physics, under Award Number DE- SC0009998. We acknowledge the use of IBM Quantum services for this work provided through Brookhaven National Laboratory. The views expressed are those of the authors and do not reflect the official policy or position of IBM or the IBM Quantum team.

\appendix
\section{Details of the variational algorithm~\label{app:var}}

The construction of thermal states of a given 
quantum many-body system is a challenging problem
for quantum computers, since a thermal state is a mixed state that cannot be obtained using only unitary quantum circuits. To find a suitable approximation to the thermal state, we note that it is possible to represent its density matrix as a convex sum\footnote{Set of operators $\rho_{i}$ such that 
$\rho = \sum_{i} p_{i} \rho_{i}$ where 
$0 \le p_{i} \le 1$ and $\sum_{i} p_{i} = 1$.}
of pure states. It can be shown that this representation is not unique. Once we find $\rho$, 
the expectation value of thermodynamic quantities
in thermal equilibrium can be computed. 
Note that other mixed states, such as incoherent mixtures, do not necessarily have a pure 
state decomposition. In that case, the procedure we use in this work is not applicable. An incoherent mixture of states is usually attributed to dissipative systems, which include thermal fluctuations and spontaneous emission. 

\begin{table}[htbp]
    \renewcommand{\arraystretch}{1.05}
\setlength{\tabcolsep}{7pt}
    \centering
    \begin{tabular}{|c|c|c|c|c|}
        \hline
        $\beta$ & $\braket{l}_{\text{VQC1,d}}$ & $\braket{l}_{\text{VQC2,d}}$ & $\braket{l}_{\text{VQC1,s}}$ & $\braket{l}_{\text{VQC2,s}}$\\
        \hline
        1 & 1 & 1 & 1 & 1 \\
        5.25 & 1.6 & 2 & 5.8 & 7.6 \\
        9.5 & 6.8 & 1 & 8.4 & 2.2 \\
        13.75 & 4.2 & 1 & 8.6 & 1.4 \\
        18 & 5.8 & 1 & 7.8 & 1.4 \\
        22.25 & 5.8 & 1 & 9.0 & 1.0 \\
        26.5 & 4.6 & 1 & 8 & 1.2 \\
        30.75 & 7.0 & 1.2 & 8 & 1.2 \\
        35 & 5.4 & 1.0 & 8.6 & 1.2 \\
        \hline
    \end{tabular}
        \caption{\label{tab:av_layers}\it Average number of layers $\braket{l}$ for each value of inverse temperature $\beta$ for the dense SYK (second and third column) and sparse SYK models (last two columns) to attain at least $90\%$ fidelity for $N=12$.}
\end{table}

\subsection{Choice of the cost function}

In the main text, we used the free energy as the cost function to construct an approximation to the thermal state of the system at some inverse temperature $\beta$. This can be seen by considering the relative entropy between two density matrices, $\rho_{p}$ and $\rho_{\beta}$, where the former denotes the approximation we prepared and the latter the true thermal density matrix. The variationally prepared thermal state $\rho_{p}$ is close to the actual thermal state if the relative entropy $S(\rho_{p}\vert \rho_{\beta}$) is close to zero. The relative entropy is minimized when $F_{p}$, i.e., the prepared free energy, is equal to the actual free energy.

\begin{table}[t] 
    \centering
    \begin{tabular}{c||c|c}
        & Dense SYK & Sparse SYK\\\hline
        $N$ & Params. per layer & Params. per layer\\\hline\hline
        6 & 15 & 15  \\
        8 & 70 &  29 \\
        10 & 210 & 45  \\
        12 & 495 & 70  \\\hline
    \end{tabular}
    \caption{\it Average number of trainable parameters per layer in each $\left\{\text{VQC}_2\right\}$ circuit. The number of trainable parameters in $\text{VQC}_1$ is $3 n \cdot l/2$ where $l$ is given in Table \ref{tab:av_layers} and $n$ is the number of qubits. Note that the number of parameters is the number of terms in the Hamiltonian.}
    \label{tab:params_all}
\end{table}

\begin{figure*}
    \centering
    \subfigure[\it Dense SYK model]{\includegraphics[width=0.80\linewidth]{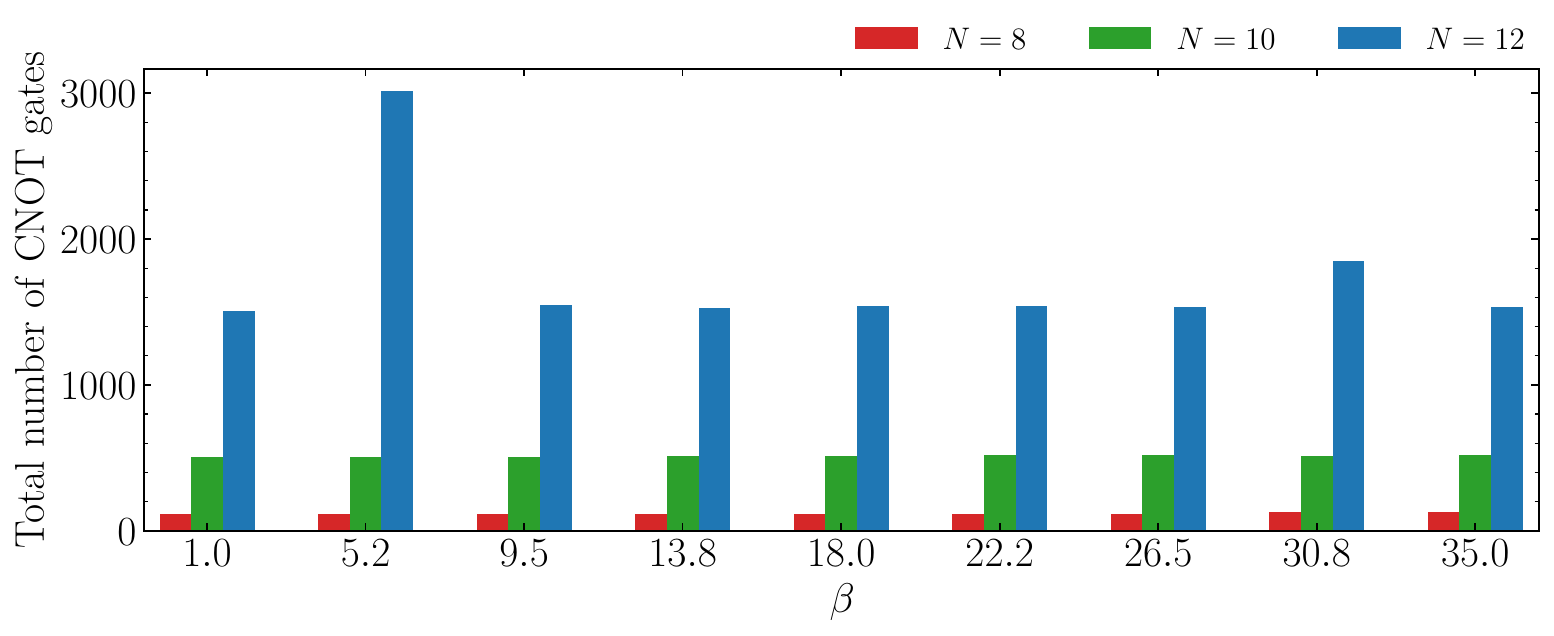}}
     \subfigure[\it Sparse SYK model]{\includegraphics[width=0.801\linewidth]{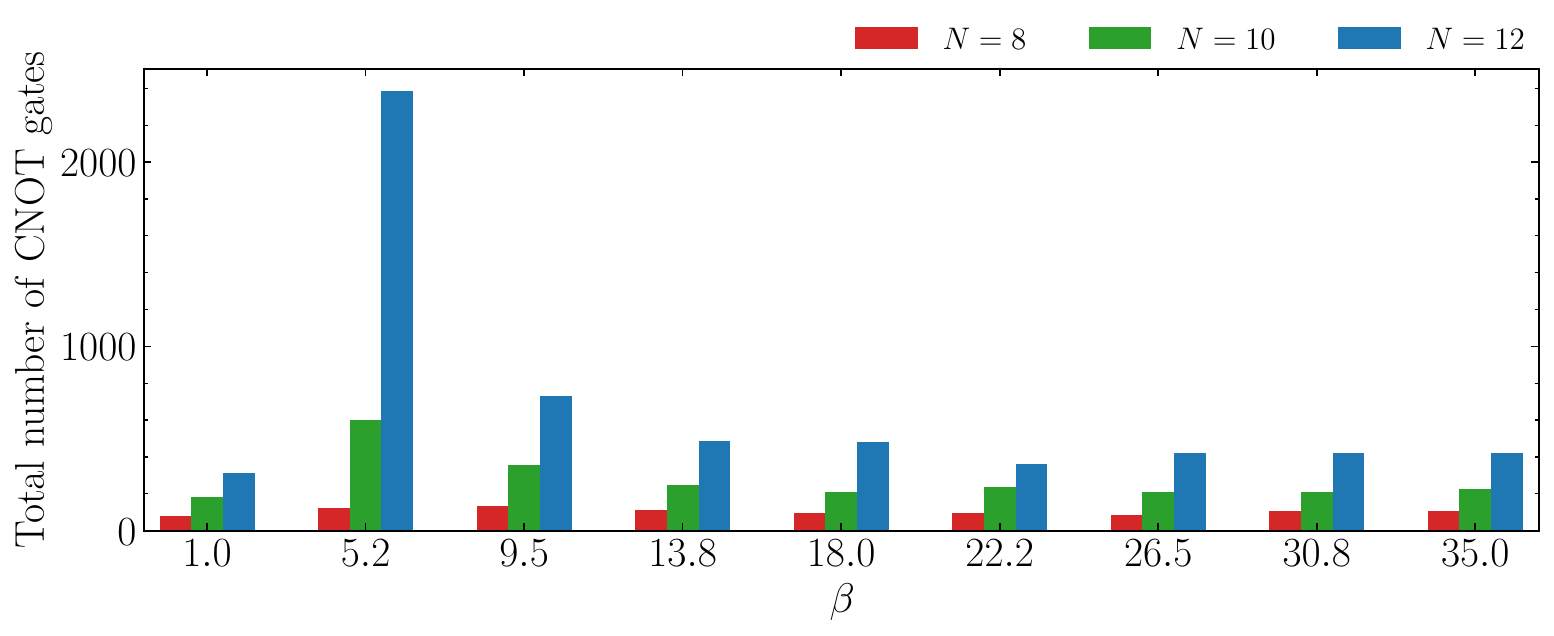}}
    \caption{\it The average total number of CNOT gates used for the thermal state preparation per instance of the a) full (or dense) and b) sparse SYK model for different $N$ and $\beta$ values. Each circuit is required to reach 90\% fidelity or a maximum of ten layers. The maximum average circuit depth in terms of two-qubit gates is $\sim 75\%$ of the total number of CNOT gates.}
    \label{fig:CNOTcount}
\end{figure*}

\CHANGE{For completeness, we derive these results. First, we show that relative entropy is lower bounded by zero. Then, we relate it to the prepared and exact free energy. 
To show the first result, consider two density matrices given by $\rho = \sum_{i} p_{i} \ket{i} \bra{i}$ and 
$\rho^{\prime} = \sum_{j} q_{j} \ket{j} \bra{j}$ respectively. We find
\begin{align}
    S(\rho\vert\rho^{\prime}) &= \rm{Tr}(\rho \log \rho) - \rm{Tr}(\rho \log \rho^{\prime})  \nonumber \\
    & = \sum_{i} p_{i} \log p_i - \sum_{i} \langle i \vert \rho \log\rho^{\prime} \vert i \rangle \nonumber \\
    & = \sum_{i} p_{i} \log p_i - \sum_{i} \sum_{j} p_{i} \langle i \vert j \rangle 
    \langle j \vert i \rangle \log q_{j} \nonumber \\ 
    & = \sum_{i} p_{i} \Big(\log p_i - \sum_{j} \langle i \vert j \rangle 
    \langle j \vert i \rangle \log q_{j} \Big)
    \label{eq:Sreldef}
\end{align}
Since the logarithm is a concave function, we have $\log(a_i) \ge \sum_{j} \langle i \vert j \rangle \langle j \vert i \rangle \log(q_j)$ where $a_i = \sum_{j} \langle i \vert j \rangle \langle j \vert i \rangle q_{j}$. Using this in~\eqref{eq:Sreldef}, we find
\begin{align}
   S(\rho\vert\rho^{\prime}) &\ge \sum_{i} p_{i} \Big(\log p_i - \log a_{i} \Big) \nonumber \\ 
   &\ge -\sum_{i} p_{i} \log\Big(\frac{a_i}{p_i}\Big)\,. 
\end{align}
Using $1-x \le -\log(x)$ for $x>0$, we obtain
\begin{align}
   S(\rho\vert\rho^{\prime}) &\ge -\sum_{i} p_{i} \log\Big(\frac{a_i}{p_i}\Big) \nonumber \\
   & \ge -\sum_{i} p_{i}\Big(1 - \frac{a_i}{p_i}\Big)  \nonumber \\
   & \ge 0.  
\end{align}
With this result that the relative entropy (also known as `relative entropy of entanglement') is non-negative, we can express it in terms of thermodynamical observables and find: 
\begin{align}
    S(\rho_{p} \vert \rho_{\beta}) &= {\rm Tr} [\rho_{p} \log(\rho_p)] - {\rm Tr} [\rho_{p} \log(\rho_\beta)] \nonumber  \\ 
    & = -S_{p} + \beta E_{p} + \log(Z_{\beta})  \nonumber \\ 
    & = - S_{p} + \beta E_{p} - \beta F_{\beta} \nonumber \\ 
    & = \beta(F_{p} - F_{\beta}) \\
    & \ge 0 \,.
\end{align}
Note that $F_{p} \ge F_{\beta}$ is sometimes referred to as the Gibbs-Bogoliubov inequality. This inequality states that at any finite temperature $T$, the free energy of the system is always smaller than the one calculated using some trial state. This inequality is the finite-temperature version of the Rayleigh-Ritz result. To construct the thermal state, one can also maximize the von Neumann entropy. However, it is computationally easier to minimize the free energy, as done in this work.} 

\subsection{Resource Estimate}

The resources required to attain a fixed fidelity depend on several factors. Table~\ref{tab:av_layers} shows the average number of layers for each $\beta$ for the first and second variational circuits for both the sparse and dense SYK models. The two-qubit cost of the second variational circuit dominates the overall cost. We show the average two-qubit gate costs for different $N$ for the models considered in Fig.~\ref{fig:CNOTcount}. 

In addition to the number of layers and associated gate counts, the optimization also depends strongly on the number of parameters, which is given by the number of single-qubit rotation gates. The count per layer of VQC2 is shown in Table.~\ref{tab:params_all}. 

\section{Thermofield double states~\label{app:TFD}}

In this section, we discuss an alternative approach to preparing the thermal state. The density matrix describing the thermal state is denoted by $\rho_A$, and the size of the associated Hilbert space is ${\rm dim}(\mathcal{H}_{A})$. The thermal state can always be purified; a procedure often referred to as the ``church of the larger Hilbert Space''~\cite{Gottesmann2001} where every mixed state is represented by a pure entangled state of a larger system. In the context of SYK-type models, these states are related to eternal black holes or wormholes and are referred to as TFD states. They play an important role in the holographic duality relating a quantum field theory in $d$ dimensions to a gravitational theory in $d+1$ dimensions. These pure states are non-unique. They can be obtained by preparing the ground state of a Hamiltonian acting on some enlarged Hilbert space, which we denote by $\mathcal{H}_{A} \otimes \mathcal{H}_{B}$. The thermal density matrix can be obtained by tracing out either subsystem $A$ or $B$. Therefore, rather than variationally constructing the thermal state directly, as described in the main text of this work, we can also construct the thermal state by first constructing the TFD states using $N$ qubits and then tracing out one of the subsystems. The TFD state is given by:
\begin{align} 
\vert {\text {TFD}}\rangle  = \frac{1}{\sqrt{Z}} \sum_i e^{-\beta E_i / 2} |i\rangle _A \otimes |i\rangle _B \,,
\end{align}
where $A$ and $B$ denote the two subsystems. Several proposals in Refs.~\cite{Maldacena:2018lmt, Cottrell:2018ash, Wu:2018nrn} have been put forth to construct the TFD state. The main idea of these constructions is that the TFD state is the ground state of a modified Hamiltonian acting on the enlarged Hilbert space. This direction has already been explored in Refs.~\cite{Zhu:2019bri, Su:2020zgc}.
\begin{figure}
    \centering
    \includegraphics[width=\linewidth]{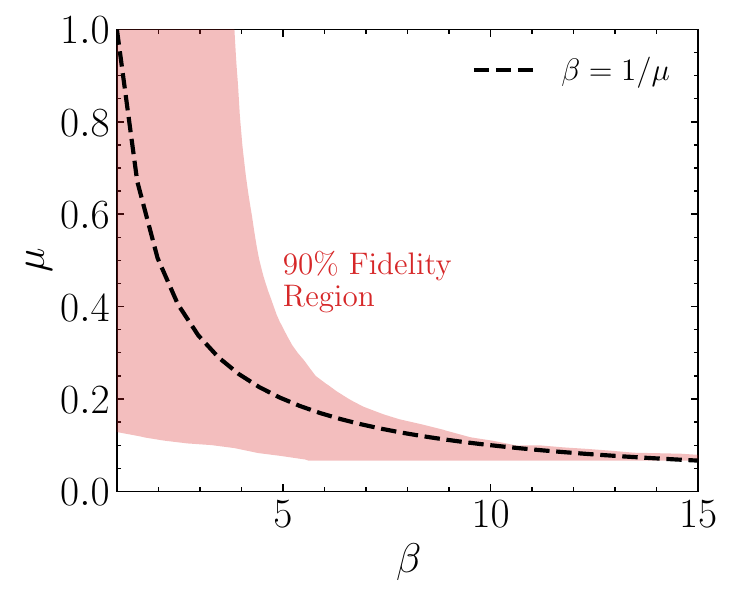}
    \caption{\it The region of $90\%$ fidelity between the thermal state and traced out TFD state for a single instance of the $N=8$ SYK model. The region covers the expected dependence of $\mu$ in the TFD Hamiltonian and $\beta$ of the thermal state. This is an additional check to verify that the construction of the thermal state can also be achieved via the TFD state construction. However, note that fewer resources are required for the former.}
    \label{fig:TFD_Gibbs_map}
\end{figure}

The Hamiltonian to construct the TFD state is given by~\cite{Maldacena:2018lmt, Cottrell:2018ash}
\begin{equation}
    H = H_{A} \otimes \mathbb{I}_B + \mathbb{I}_A \otimes H_{B} + H_{AB}\,.
    \label{eq:TFD_ham} 
\end{equation}
Here $H$ is an $N$-qubit Hamiltonian, and $H_{A,B}$ act on the $N/2$-qubit subsystems. For our case, $H_{A}$ and $H_{B}$ are SYK Hamiltonians with the same disorder couplings. The proposal given in Ref.~\cite{Maldacena:2018lmt}, which was also studied in Ref.~\cite{Su:2020zgc}, is to consider interactions between the two subsystems given by
\begin{equation}
    H_{AB} = i \mu \sum_{j=1}^{N} \chi^{j}_{A} \otimes \chi^{j}_{B}\,.
    \label{eq:TFD_hamHAB}
\end{equation} 
If we construct the ground state of the Hamiltonian in Eq.~\eqref{eq:TFD_ham} and trace out either subsystem, we should obtain the corresponding thermal state at some inverse temperature $\beta$ for the other subsystem. This can be seen by considering the density matrix corresponding to the TFD state $\rho_{AB} = \ket{\text{TFD}} \bra{\text{TFD}}$ and tracing out the subsystem $B$. We find
\begin{align}
    \rho_{A} &= \text{Tr}_{B}[\rho_{AB}] \nonumber \\
    &= \frac{1}{Z} \sum_{j} \langle j \vert 
    \sum_{i,i^{\prime}}  e^{-\frac{\beta E_{i}}{2}} \ket{i}_{A} \ket{i}_{B} \bra{i^{\prime}}_{B} \bra{i^{\prime}}_{A} e^{-\frac{\beta E_{i{\prime}}}{2}} \vert j \rangle \nonumber  \\
    &=  \frac{1}{Z} \sum_{i} e^{-\beta E_{i}} \ket{i}_{A} \bra{i}_{A} \nonumber  \\ 
    & = \frac{e^{-\beta H_{A}}}{Z} \,.
\end{align}
We prepare the ground state of the Hamiltonian in Eq.~(\ref{eq:TFD_ham}) using a variational algorithm. After tracing out subsystem $B$, we compare our result to the thermal state using exact diagonalization. The numerical results are shown in  Fig.~\ref{fig:TFD_Gibbs_map}, where we highlight the region where the two states agree with a fidelity of at least $90\%$. As expected, this region includes $T = \mu$. Note that here, we consider a single instance of the dense SYK model for $N=8$. The figure shows that the energy expectation value 
$E(\mu)$ obtained from $\rho_{AB}$ is related to 
$E(\beta)$ obtained from $\rho_{A}$ by a well-defined relation
for large $\mu$. The dependence becomes sublinear at small values of $\mu$ (not shown). This has been studied in detail in Refs.~\cite{Maldacena:2016hyu, Lantagne-Hurtubise:2019svg}. 

\bibliographystyle{utphys}
\raggedright
\bibliography{v1.bib}

\end{document}